# Statistical multiscale mapping of IDH1, MGMT, and microvascular proliferation in human brain tumors from multiparametric MR and spatially-registered core biopsy

Jason G Parker[1,2], PhD, Emily E Diller[2,1], MS, Sha Cao[3], PhD, Jeremy T Nelson[4,1], PhD, Kristen Yeom[5], MD, Chang Ho[1], MD, Robert Lober[6], MD, PhD

[1]Department of Radiology and Imaging Sciences, Indiana University School of Medicine; [2]School of Health Sciences, Purdue University; [3]Department of Biostatistics, Indiana University School of Medicine; [4]Military Health Institute, University of Texas Health San Antonio; [5]Radiology, Lucile Salter Packard Children's Hospital and Stanford University Medical Center; [6]Neurosurgery, Dayton Children's Hospital

<u>Corresponding Author Info</u>: Jason Parker, 950 W. Walnut St., Indianapolis, IN 46202, telephone: (317) 274-2072, fax: (317) 274-1067, parkerjg@iu.edu

## Abstract

<u>Purpose</u>: We propose a statistical multiscale mapping approach to identify microscopic and molecular heterogeneity across a tumor microenvironment using multiparametric MR (mp-MR).

<u>Methods</u>: Twenty-nine patients underwent pre-surgical mp-MR followed by MR-guided stereotactic core biopsy. The locations of the biopsy cores were identified in the pre-surgical images using stereotactic bitmaps acquired during surgery. Feature matrices mapped the multiparametric voxel values in the vicinity of the biopsy cores to the pathologic outcome variables for each patient and logistic regression tested the individual and collective predictive power of the MR contrasts. A non-parametric weighted k-nearest neighbor classifier evaluated the feature matrices in a leave-one-out cross validation design across patients. Resulting class membership probabilities were converted to chi-square statistics to develop full-brain parametric maps, implementing Gaussian random field theory to estimate inter-voxel dependencies. Corrections for family-wise error rates were performed using Benjamini-Hochberg and random field theory, and the resulting accuracies were compared.

<u>Results</u>: The combination of all five image contrasts correlated with outcome ($P<10^{-4}$) for all four microscopic variables. The probabilistic mapping method using Benjamini-Hochberg generated statistically significant results ($\alpha \leq .05$) for three of the four dependent variables: 1) IDH1, 2) MGMT, and 3) microvascular proliferation, with an average classification accuracy of 0.984 ± 0.02 and an average classification sensitivity of 1.567% ± 0.967. The images corrected by random field theory demonstrated improved classification accuracy (0.989 ± 0.008) and classification sensitivity (5.967% ± 2.857) compared with Benjamini-Hochberg.

<u>Conclusion</u>: Microscopic and molecular tumor properties can be assessed with statistical confidence across the brain from minimally-invasive, mp-MR.

## Introduction

Emerging targeted therapies interfere with specific molecules that promote tumor growth and infiltration based on patient-specific predictive cellular and molecular biomarkers [1]. However, heterogeneous genomic and phenotypic tumor microenvironments contribute to incomplete treatment by targeted therapy and promote tumor recurrence via a non-linear branched evolution of the cancer genome [2],[3]. Biopsy is currently the most effective method to assess patient-specific tumor biomarkers for targeted therapeutics, but clinical outcomes are limited by tumor heterogeneity which cannot be assessed by invasive biopsy alone [1]. Medical imaging techniques that are minimally-invasive and assess cellular and molecular tissue characteristics across the entire tumor bed and tumor microenvironment (TME) hold the potential to significantly improve the characterization and treatment of aggressive brain tumors [5]-[7].

Significant efforts are underway to develop tumor heterogeneity mapping techniques using minimally-invasive imaging including texture analysis [8]-[10]; proton [11],[12] and hyperpolarized $^{13}$C [13] spectroscopy; and most recently MR fingerprinting [14]. Generally these methods classify tumor properties at one of two levels: 1) volumetrically, by segmenting adjacent voxels together into classes, or 2) on a voxel-wise basis, treating each voxel independently. Volumetric segmentation techniques leverage spatial correlations in adjacent voxels that may be associated with tumor biology and/or the physical attributes of the acquisition process to improve SNR and classification accuracy. However, these improvements are balanced by a decrease in the theoretical spatial resolution of the parametric images, ultimately limiting the assessment of heterogeneity. Voxel-wise methods have a theoretical spatial resolution on the order of a single voxel, but suffer from significantly increased noise, which may be counteracted by the concomitant acquisition of multiple MR signatures. A recent voxel-wise algorithm demonstrated the ability to map tumor cellularity from three MR contrasts when biopsy findings were localized to the pre-surgical images [15]. Functional Diffusion Maps (fDMs) have also been estimated from ADC maps by identifying biopsy core locations on intra-operative computed tomography and post-surgical high resolution 3D anatomical images [16]. Alternatively, MR Fingerprinting (MRF) is a promising voxel-wise approach that has been successfully used to parameterize important tumor tissue properties including T1, T2, and M0, as well as physical system properties including B0 and B1 [17]. There is some emerging evidence that MRF can be used to map functional tissue parameters including perfusion, oxygenation, and microvascular structure [18], but the extent to which the MRF technique can be applied to functional, cellular, and molecular imaging remains unknown.

Here we propose to map cellular and molecular tumor properties throughout the TME in a voxel-wise manner by leveraging the growing dimensionality of clinical MR data. Our approach does not inherently rely on spatial correlation information or simulations of various tissue properties for classification. Instead, we hypothesize that the dimensionality of MR data alone provides a readily available vehicle to traverse tissue scale. We evaluate our hypothesis in three separate sub-steps: ***Sub-hypothesis 1)*** significant relationships ($\alpha \leq 0.05$) between macro- and micro-scale properties can be identified using elementary statistical testing when surgical pathology results are localized to the pre-surgical image space; ***Sub-hypothesis 2)*** non-parametric machine learning can classify microscopic properties from macroscopic images with high accuracy ($\geq 95\%$) when traditional corrections for family-wise error rates are employed; and ***Sub-hypothesis 3)*** clinically-useful multiscale classification across the entire image space can be accomplished when the parametric images are treated as Gaussian random fields.

Experimentally, we developed a data-driven model linking spatially registered core biopsy data to multiparametric MR. We used a diverse patient population consisting of more than 10 different disease classes, making the microscopic classifications more difficult but also more generalizable to a clinical population. We performed initial statistical evaluations on the model to determine the feasibility of predicting the biopsy findings from the MR values alone. We then evaluated the use of non-parametric machine learning to predict four clinically relevant properties: IDH1 mutation status, MGMT promoter methylation, cellular necrosis, and microvascular proliferation. Class membership probabilities output from the machine learning model were converted to chi-square statistical estimates using probabilistic distributions of the dependent variables identified *a priori*. The Benjamini-Hochberg algorithm controlled for family-wise error rates (FWER), and the classification accuracy and sensitivity of the results were optimized across a single classification tuning parameter. Finally, the machine learning model was extended to calculate chi-square ($\chi^2$) parametric maps across the entire brain of all 29 patients. To improve statistical classification sensitivity in the image domain, we implemented Gaussian random field theory (RFT) to estimate the interdependence of voxels and then group statistical findings into thresholded clusters. We evaluated the images qualitatively by clinical experts and quantitatively by classification accuracy in the biopsy sample volume.

## **Methods**

### *Study population and model development*

The Indiana University Institutional Review Board (IRB) approved this retrospective study without patient consent under the conditions that all patient data would be de-identified upon the completion of patient enrollment. De-identification consisted of removing all 18 HIPAA Privacy Rule identifiers from images, pathology reports, and clinical data. Accordingly, all dates were removed, but age and the difference in days between imaging and biopsy were retained for each patient. This study was not listed on ClinicalTrials.gov, and no part of the dataset presented here has been used or published on in the past. All source data used in this paper are openly shared with the radiology community for research replication and further analyses at http://www.iu.edu/~mipl. Inclusion criteria for this study required that the patients 1) had previously undergone targeted (stereotactic image-guided) core biopsy of the brain at our institution with at least three orthogonal plane images saved showing the location of the core; 2) had completed an MR scan a maximum of 60 days prior to biopsy that included at least T1 weighting ($T1_w$), T1 weighting post gadolinium injection ($T1_{w-post}$), T2 weighting ($T2_w$), T2 weighting with fluid attenuated inversion recovery (T2-FLAIR), and diffusion weighted imaging (DWI). We specifically did not limit the study to a single tumor type (e.g. glioma) to ensure that the non-parametric model could be tested in a clinically-relevant population. Approximately 100 patients were screened, and 29 met the criteria for enrollment (N=29). The characteristics of the enrolled population are shown in **Table I**. All pathology reads and diagnoses were performed by two experienced neuropathologists, each with more than 10 years'

**Table I.** Subject population characteristics

| Parameter | Value |
|---|---|
| N | 29 |
| Sex | |
|   F | 13/29 (45%) |
|   M | 16/29 (55%) |
| Age (y) | |
|   Mean ± standard deviation | 56.4 ± 19.3 |
|   Range | 23-89 |
| Pathology-based diagnosis | 29 |
|   Glioma | 16 |
|     WHO Grade IV | 7 |
|     WHO Grade III | 2 |
|     WHO Grade II | 3 |
|     WHO Grade I | 4 |
|   Metastatic carcinoma | 4 |
|     Breast | 2 |
|     Lung | 1 |
|     Melanoma | 1 |
|   Diffuse large B-cell lymphoma | 2 |
|   Schwannoma | 1 |
|   Reactive changes | 2 |
|   Abscess | 1 |
|   Germinoma | 1 |
|   Demyelination | 1 |
|   Normal | 1 |
| Time between biopsy and imaging (days) | |
|   Mean ± standard deviation | 9.7 ± 9.1 |
|   Range | 0-37 |
| Biopsy samples | |
|   Mean number of samples taken per patient ± standard deviation | 3.24 ± 1.8 |
|   Mean sample volume ($mm^3$) | 4119.5 |

*Note – Unless otherwise noted data are specified as number of patients.

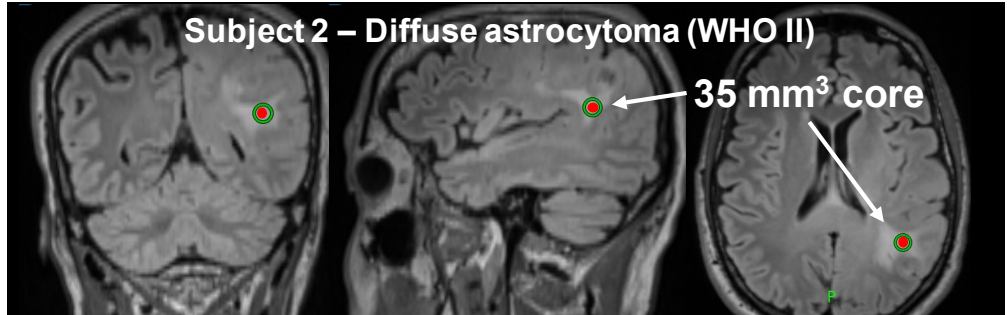

**Figure 1.** Example neuronavigation targeting images for Subject 2.

experience practicing in an academic medical center.

The five MR sequences were the only independent variables used in this analysis – for an overview of the acquisition parameters please see *Supporting Information – Supplemental Table I*. Approximately 90% of the acquisitions were performed at 1.5T (26 of 29), and approximately 70% of the anatomical sequences used 3D readout (101 of 145). All DWI acquisitions used two b-values (0,1000 s/mm$^2$) and 3 orthogonal directions. For post-processing, all images were initially registered to the $T1_{w-post}$ frame-of-reference for each patient. $T1_w$ was registered using a 12 degree-of-freedom (DOF) transform and minimization of a correlation ratio objective function [19]. $T2_w$, T2-FLAIR, and DWI (B0-only) were registered using a 12 DOF transform and minimization of a mutual information objective function [20]. Apparent diffusion coefficient (ADC) maps were then registered to the individual $T1_{w-post}$ reference frame by applying the affine transformation matrix estimated for the DWI B0 images. We normalized voxels of each contrast to the mean value of uninvolved white matter determined by a spherical region-of-interest on the $T1_{w-post}$. The 3-dimensional centroid of the biopsy core was identified on the $T1_{w-post}$ for each patient by visually comparing the three-plane neuronavigation plans (**Figure 1**) to the pre-intervention MR images. We used the size of the biopsy core as reported in the pathology report to define a sphere centered at the location of the biopsy needle tip from which the image contrasts were extracted. This method ensured the feature matrix and subsequent machine learning model included only those voxels representative of biopsied tissues.

We extracted the four dependent categorical variables from clinical pathology reports for each patient, classifying voxels as IDH1 mutation status positive ($IDH1_{MS+}$) if the corresponding specimen contained any IDH1-R132H-positive cells based on immunohistochemistry, and voxels as MGMT promoter methylation status positive ($MGMT_{PMS+}$) if present based on a methylation-specific PCR-based assay. When applicable, a clinical pathologist evaluated several representative microscopic sections for the presence of cellular necrosis ($CNEC_+$) and/or microvascular proliferation ($MVP_+$). Because this study was not limited to primary brain tumors, the pathologist used their discretion to determine which tests should be applied on an individual patient basis, as a standard of care. Importantly, if the physician determined a variable need not be measured for a given patient, we classified it as negative for the analysis.

A biostatistician and co-author on this paper (S.C.) guided, oversaw, and reviewed the statistical analyses; an overview is given in **Figure 2**.

*Sub-hypothesis 1: Elementary statistical evaluations*

First, we calculated the normalized contrast values for the independent variables across the entire biopsy sphere for each patient, and combined them into a single feature vector mapping the five independent variables to the four outcome variables for each voxel. This resulted in a feature matrix of size 147,031 rows x 9 (binary) columns. A binary logistic regression was performed for each dependant variable by fitting a maximum-likelihood logit model. The regressions were sample weighted by the inverse of the probability of inclusion due to the sampling design to account for class imbalances [21]. The results characterized the overall (combined) predictive power of the five image contrasts for each microscopic variable using the Wald $\chi^2$ test and McFadden's pseudo $R^2$ [22].

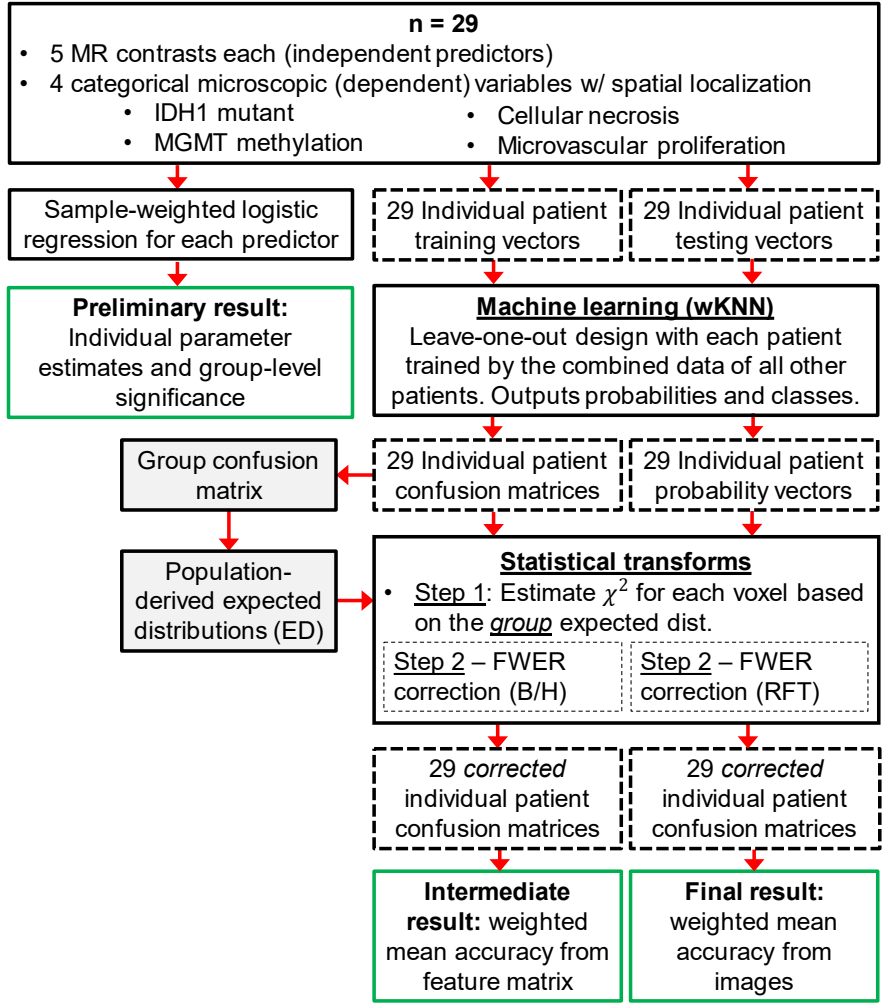

**Figure 2.** Flowchart of the processing and statistical analysis steps. Endpoints resulting in statistical conclusions are outlined in green.

*Sub-hypothesis 2: Multiscale classification without spatial information*

Next, we developed individual training feature matrices ($FV_{i,train}$) and testing feature matrices ($FV_{i,test}$) for each patient, $i$. The $FV_{i,train}$ and $FV_{i,test}$ matrices included the following data: each of the 5 normalized MR contrast values in columns 1-5; the subject number ($i$) in column 6; and the binary class flag for $IDH1_{MS+}$, $MGMT_{PMS+}$, $CNEC_+$, and $MVP_+$ in columns 7-10, respectively. The rows of $FV_{i,train}$ corresponded to the biopsied voxels across all patients except patient $i$; the rows of $FV_{i,test}$ corresponded to the biopsied voxels for patient $i$. A set of 116 machine learning experiments (29 patients x 4 dependent variables) were then carried out using a leave-one-out design to ensure that in no case could data from the same patient be used for both training and testing.

The machine learning classifier was a non-parametric weighted k-nearest neighbor design (wKNN) [23] with class weights calculated by the inverse Euclidean distance. The only tuning parameter used for classification was the number of neighbors, $k$, included in the class calculations. The classifier output was a 2-element vector for each voxel representing the probability of membership in each binary class, calculated as the normalized sum of the inverse Euclidean distance. We transformed the probability vectors across the biopsy volume for each patient to a chi-square test statistic ($\chi^2$) using Pearson's method [24]. The statistic compared our predicted class probability for each voxel with the background probability calculated for the entire voxel population across all patients. The chi-square transform was chosen (i.e. instead of z or t distributions) because the background probabilities could be explicitly calculated from the data. A clinical implementation of this algorithm would similarly have access to background population probabilities assuming the availability of a robust training dataset. The $\chi^2$ values were then thresholded to a given $\alpha$-value using standard statistical transforms. For FWER correction, a $p$-value threshold was calculated for each patient using the Benjamini-Hochberg procedure [25] at an $\alpha$ of 0.05. We calculated a confusion matrix for each patient by choosing the class of greatest probability for all voxels that passed the FWER correction. The final measure of classification accuracy, $ACC(k)$, was calculated as the mean accuracy across all 29 confusion matrices, with optimization across the tuning parameter $k$. The final measure of classification sensitivity, $SENS(k)$, was calculated as the percent of voxels sampled by biopsy that met the $\alpha$ threshold.

*Sub-hypothesis 3: Multiscale classification across the image space*
$\chi^2$ parametric maps of each microscopic variable were then calculated as before for every voxel and overlayed on the T1$_{w-post}$ images using the tuning parameter that yielded the greatest value of $SENS(k)$ at an $ACC(k) \geq 0.95$. Because the images resulted in several orders of magnitude more voxels to be classified than in any of the $FV_{i,test}$ vectors, we determined that a less conservative FWER correction approach was necessary. As the $\chi^2$ maps were smooth statistical fields, we used a mature FWER correction technique widely used in functional MRI which first estimates the spatial correlation of the statistical image and then identifies clusters of voxels which result in the expected Euler characteristic (EC) for a smooth statistical map [26]-[28]. We performed both the spatial correlation and EC optimizations using FSL [29] [30], resulting in $\chi^2$ parametric images for each dependent variable that controlled FWER at the 5% level.

**Results**

*Study population and model development*
The enrolled population had a median age of 59 years (max 89, min 23) and had 16 males (55%). The mean difference in time between imaging and biopsy was 9.7 ± 9.1 days. The biopsy-confirmed diagnoses included: sixteen gliomas (seven Grade IV, two Grade III, three Grade II, and four Grade I), four metastatic carcinomas (two breast, one lung, and one melanoma), two diffuse large B-cell lymphoma, one schwannoma, two reactive changes, one abscess, one germinoma, one demyelination, and one normal. A detailed overview of the enrolled subject population and demographics is given in **Table I**.

*Sub-hypothesis 1: Elementary statistical evaluations*
The combination of all five image contrasts was found to be significantly correlated with outcome ($P < 10^{-4}$) for all four microscopic variables **(Table II)**. IDH1$_{MS+}$ had the greatest likelihood with a pseudo $R^2$ of 0.26, followed by $MGMT_{PMS+}$ (0.25), $CNEC_+$ (0.22), and $MVP_+$ (0.07). For a complete breakdown of the prediction results by individual image contrast, please see *Supporting Information – Supplemental Table II.*

**Table II.** Overall prediction results of the combined (5) image contrasts from binary logistic regression analyses.

| Outcome | Observations | Wald $\chi^2$ | P > $\chi^2$ | Pseudo R² |
|---|---|---|---|---|
| IDH1$_{MS+}$ | 147,031 | 18,520.37 | 10$^{-4}$ | 0.2637 |
| MGMT$_{PMS+}$ | 147,031 | 8437.73 | 10$^{-4}$ | 0.2455 |
| CNEC$_+$ | 147,031 | 22,245.20 | 10$^{-4}$ | 0.2180 |
| MVP$_+$ | 147,031 | 17,136.48 | 10$^{-4}$ | 0.0661 |

**Figure 3** shows the parameter estimates (regression coefficients), demonstrating that characteristic patterns of the logits across the five predictors exist for each microscopic variable, even when accounting for the robust standard errors. Of note, $IDH1_{MS+}$ and $MGMT_{PMS+}$ exhibited strong negative correlations with $T1_w$, and $MGMT_{PMS+}$ also displayed a large negative correlation with ADC. $CNEC_+$ demonstrated a strong positive correlation with $T1_w$, while $IDH1_{MS+}$ has a strong positive relationship with T2-FLAIR. These initial findings provided a statistical foundation upon which our hypothesis could then be tested using the previously described machine learning techniques.

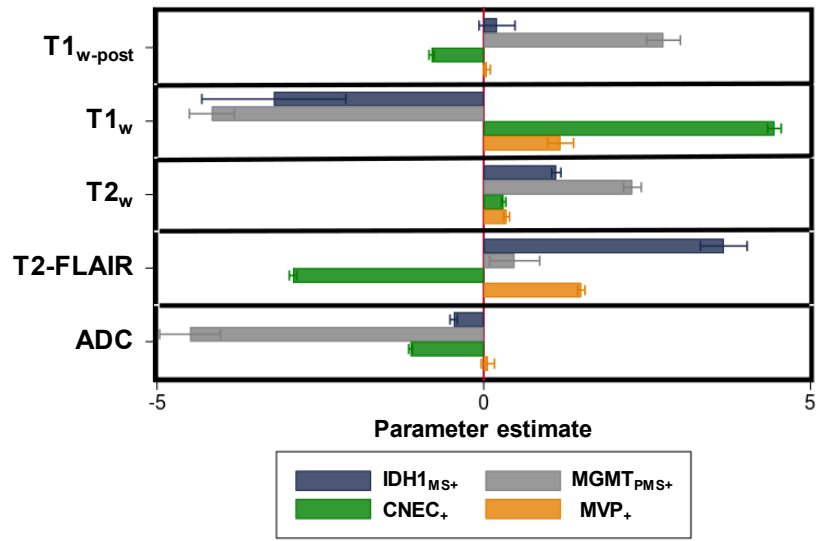

**Figure 3.** Regression coefficients for each microscopic variable across the 5 image contrasts. Error bars represent robust standard errors.

*Sub-hypothesis 2: Multiscale classification without spatial information*

The results of the machine learning optimization procedure are shown in **Figure 4**. Accuracy and the number of statistically significant voxels are shown in black and blue, respectively. The plots demonstrate that the tuning parameter $k$ has a large effect on the number of voxels which pass the FWER correction, and thus, indirectly, the overall accuracy calculation. There was similar classification behavior between $IDH1_{MS+}$ and $MGMT_{PMS+}$, in which classification accuracy generally increased with $k$, and then plateaued as the number of significant voxels began to decrease. No voxels passed the FWER correction for cellular necrosis at any value of $k$ that was tested. The classification accuracy of MVP had an approximately linear relationship with $k$, while the number of voxels passing the FWER threshold had an approximately inverse linear dependence on $k$.

From the optimization plots, we chose a tuning parameter that maximized the accuracy and the number of voxels that passed the significance threshold. In keeping with our $\alpha$ threshold of 0.05, we limited the minimum acceptable classification accuracy to be 0.95; thus, the optimal tuning parameter, $k_{opt}$, was that which maximized $SENS(k)$ in the condition that $ACC(k) \geq 0.95$. **Table III** shows the optimized tuning parameter and classification results for the four microscopic variables. The values of $k_{opt}$ for each outcome are also shown as vertical green bars in **Figure 4**.

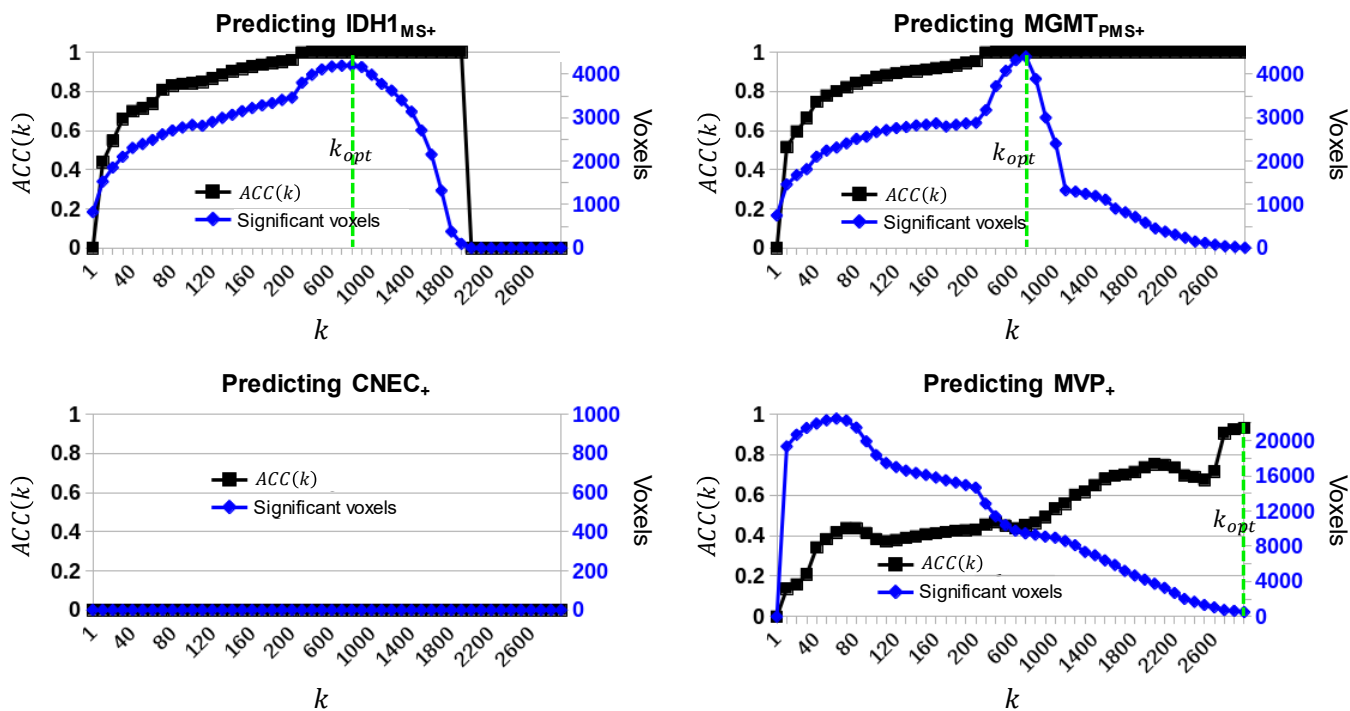

**Figure 4.** Accuracy (black; left vertical axis) and number of significant voxels (blue; right vertical axis) vs. the wKNN tuning parameter $k$. The optimal tuning parameter value ($k_{opt}$) maximized the number of significant voxels when $\alpha \leq 0.05$. $k_{opt}$ is indicated by a vertical green line for each outcome variable.

Of the 3 variables that had significant findings (IDH1, MGMT, MVP) the average classification accuracy was 0.984 ± 0.02 and the average classification sensitivity was 1.567% ± 0.967. Optimal classification results for the molecular markers IDH1$_{MS+}$ and MGMT$_{PMS+}$ were similar, both yielding an $ACC(k)$ of 1.0 and a $SENS(k)$ slightly greater than 2%. The optimal $ACC(k)$ of MVP$_+$ was 0.951 with a $SENS(k)$ of 0.2%. These results confirmed our hypothesis that multiscale classification could be performed without spatial information. However, the low number of voxels passing the correction threshold supported further evaluation of a FWER-correction technique that was more sensitive to classification.

**Table III.** Optimized results from the leave-one-out machine learning classification using Benjamini-Hochberg correction without spatial correlation information.

| Microscopic variable | $k_{opt}$ | $ACC(k)$ | # of significant voxels | $SENS(k)$ (%) |
|---|---|---|---|---|
| IDH1$_{MS+}$ | 800 | 1.0 | 4197 | 2.2 |
| MGMT$_{PMS+}$ | 700 | 1.0 | 4399 | 2.3 |
| CNEC$_+$ | N/A | 0.0 | 0 | 0 |
| MVP$_+$ | 3000 | .951 | 405 | 0.2 |

*Sub-hypothesis 3: Multiscale classification across the image space*

Example images corrected by RFT using $k_{opt}$ for IDH1$_{MS+}$, MGMT$_{PMS+}$, and MVP$_+$ are shown in **Figure 5** for 3 exemplary patients. Images for 4 additional patients are given in *Supplemental Materials – Supp. Figure I*. In **Figure 5**, the biopsy site for each patient is indicated with a yellow crosshair on the zoomed-in $\chi^2_{RFT}$ maps (2nd row), and the original uncorrected probability maps generated by the machine learning model are shown in row 3.

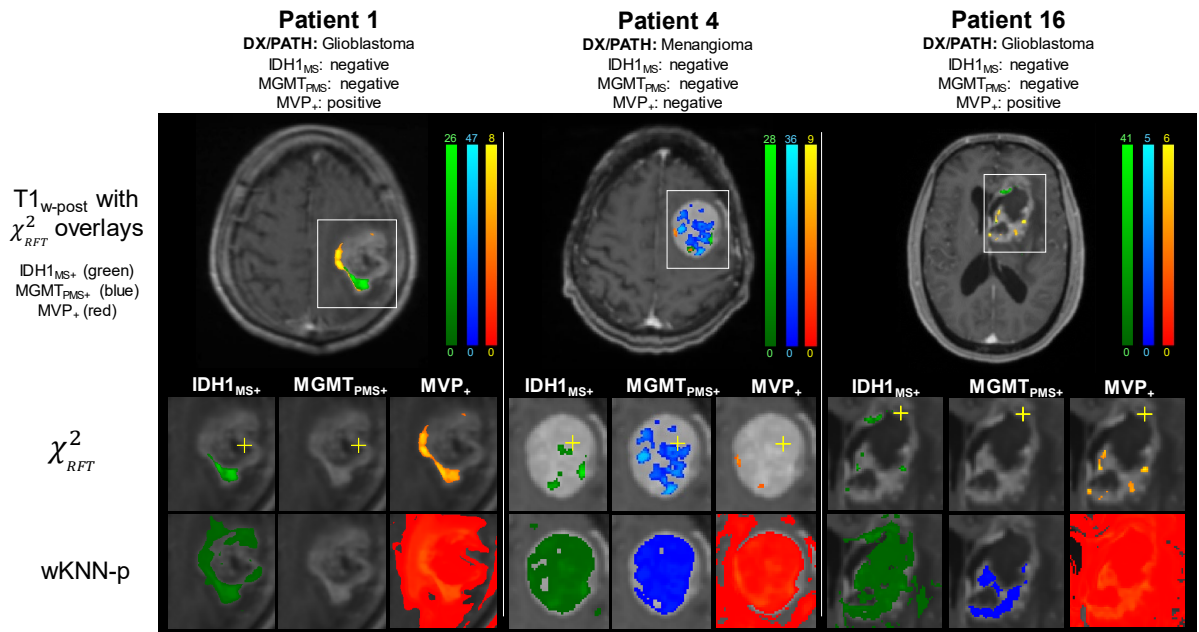

**Figure 5.** Results of the statistical mapping procedure for 3 select patients, with the location of the biopsy marked with a yellow plus sign. In all cases the $\chi^2$ image with random field theory correction dramatically reduces the number of false positive findings and demonstrates smooth noise properties across space.

Qualitatively the images demonstrate smooth statistical fields that are well localized to the tumor bed and TME. The quantitative classification results based on the $\chi^2_{RFT}$-corrected images are shown in **Table IV**. RFT demonstrated improved average classification accuracy (0.989 ± 0.008) and sensitivity (5.967% ± 2.857) compared with Benjamini-Hochberg. Notably, $SENS(k)$ for MVP₊ increased to 9.9% using RFT compared with 0.2% with Benjamini-Hochberg.

**Table IV.** Optimized results from the leave-one-out machine learning classification using random field theory correction including spatial correlation information.

| Microscopic variable | $k_{opt}$ | $ACC(k)$ | # of significant voxels | $SENS(k)$ (%) |
|---|---|---|---|---|
| IDH1$_{MS+}$ | 800 | 1.0 | 4732 | 3.2 |
| MGMT$_{PMS+}$ | 700 | 0.987 | 7096 | 4.8 |
| CNEC₊ | N/A | 0.0 | 0 | 0 |
| MVP₊ | 3000 | .982 | 14610 | 9.9 |

**Figure 6** shows an example GBM subject had significant results for all 3 outcome variables that nearly covered the entire TME (Subject 8). The 5 predictor contrasts are shown along the left side of the image zoomed-in on the tumor bed and TME. The colormaps are windowed from 0 to the maximum $\chi^2_{RFT}$ statistic (right side colorbars). Potential significant findings for MVP₊ outside the T2-FLAIR abnormality (3 red speckles frontal and medial to the tumor) may hold important information related to microscopic disease spread.

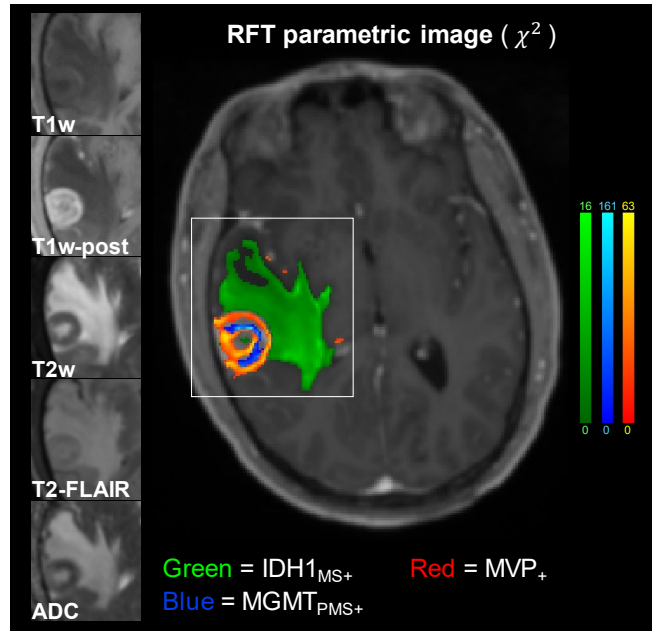

**Figure 6.** Extensive visualization of statistical confidence ROI's mapping genomic and cellular heterogeneity in a GBM patient.

## Discussion and Conclusions

Cellular and molecular heterogeneity is a significant driver of brain tumor morbidity that cannot be assessed by biopsy alone. This paper demonstrated three different methods to predict microscopic cellular and molecular properties of brain tumors from macroscopic, minimally-invasive clinical images. Elementary statistical evaluations demonstrated that significant relationships between the macroscopic and microscopic variables of interest did exist. Machine learning combined with a conservative correction for family-wise error rates was able to predict cellular and molecular properties with high accuracy but limited classification sensitivity (0.2-2.3%) for three of the four outcome variables. When spatial correlations across voxels were taken into account using Gaussian random field theory, high accuracy was retained with a significant increase in classification sensitivity (3.2-9.9%). The images generated by random field theory demonstrated acceptable noise and spatial resolution properties for clinical interpretation. Taken together, our results show that *in vivo* microscopic and even genomic mapping of human brain tumors may be clinically possible in the near future.

The near-term implication of our findings is that researchers and clinicians utilizing machine learning to predict tumor heterogeneity should consider dimensionality to be one potential vehicle by which *in vivo* imaging signatures may be used to traverse scale. The rapid expansion of anatomical and functional MR sequences and the growing availability of hybrid imaging systems only serve to enhance this opportunity. The long-term implications of our findings are that it may be possible to map cellular and molecular tumor properties across both space and time during treatment, allowing for highly personalized treatment strategies that are not currently possible. For example, MGMT promoter methylation status can vary across the tumor bed and microenvironment [31] making treatment planning challenging. Patients who are determined by surgical biopsy to have $MGMT_{PMS+}$ are expected to demonstrate good response to standard of care treatment with concomitant and adjuvant radiation therapy and chemotherapy with temozolomide [32], although this is almost always followed by relapse and eventual death. A subset of these patients are expected to also have undiagnosed $MGMT_{PMS-}$ properties, and thus may benefit from experimental personalized therapies [33]. The ability to map MGMT promotor methylation status with a minimally-invasive in vivo probe would allow for better treatment selection and drug combinations than is currently possible.

This study provides initial evidence in support of our hypothesis; however, there are significant limitations on the generalizability of our findings due to our study design. First, the retrospective design used in this paper did not allow for standardization of the immunohistochemical and molecular tests used across patients. This drawback required our analysis to rely on the clinical expertise of the pathology physicians in determining which tests were required at the individual patient level. Furthermore more comprehensive genomic evaluations (i.e. genome-wide association) could have been conducted to identify other predictor-outcome relationships than the four we investigated. The MR sequence parameters used for the five predictor variables varied across patients and locations which may have diminished their individual and collective effect sizes. The number of MR sequences was limited to 5, although many other sequences could have been used including perfusion imaging, chemical exchange saturation transfer, and MR spectroscopy. Finally, although the diversity of our patient cohort was clinically relevant, it very likely weakened our control over the experimental variables and ultimately reduced our statistical effect sizes compared with a highly controlled study focused on a single tumor or tissue type. However, the ubiquitous drawback of highly-controlled, single-disease radiomics studies is a failure to generalize to a clinically-relevant patient population [34],[35].

In summary, we have demonstrated statistical relationships between routine multiparametric imaging signatures and underlying cellular and molecular properties of brain tumors. We have applied advanced statistical methods to correct for the family-wise error rate problem associated with whole-brain statistical parametric mapping, and have shown that the results have strong agreement with surgical biopsy. These results imply that cellular and molecular mapping of tumor heterogeneity from minimally-invasive images may be possible in the near future.

*Acknowledgements*: We thank Ms. Nichole Johnson for her assistance in operating the neuronavigation system, interpreting the coordinate system and uncertainty indices, and exporting the biopsy screen captures. We thank Dr. Anderson Winkler for insight into the application of random field theory to various parametric maps, and specific issues related to implementation of our method using FSL. We thank Dr. Mark Holland for useful conversations related to the use of random field theory to estimate smoothness in the chi-square maps.

## Supplementary Materials

**Supplemental Table I.** Mean ± standard deviation of sequence parameters across the 5 image contrasts. Parameters that are not specified varied too dramatically across sequence implementation (i.e. 2D vs. 3D read-out) or were not applicable to the sequence.

| Parameter | T1$_w$ | T1$_{w\text{-post}}$ | T2$_w$ | T2-FLAIR | DWI |
|---|---|---|---|---|---|
| TR (ms) | - | - | 3461 ± 955.3 | 5800 ± 1751.2 | 10208 ± 4632.9 |
| TE (ms) | 4.8 ± 3.4 | 3.9 ± 2.5 | 151.9 ± 95.8 | 330.4 ± 97.8 | 88 ± 14.3 |
| TI (ms) | - | - | - | 1937 ± 294 | - |
| Flip angle (°) | - | - | - | - | 99 ± 28.5 |
| Pixel size (mm) | 0.8 ± 0.3 | 0.9 ± 0.2 | 0.7 ± 0.2 | 1.0 ± 0.1 | 1.3 ± 0.3 |
| Bandwidth (Hz/pixel) | 173.0 ± 22.1 | 1088.1 ± 142.3 | 360.7 ± 252.5 | 569.8 ± 221.1 | 1088.1 ± 142.3 |

**Supplemental Table II.** Individual predictor results from the binary logistic regression analyses.

| Outcome | Predictor | Coefficient | Robust Std. Error | Z | P > \|Z\| | 95% Conf. Interval | |
|---|---|---|---|---|---|---|---|
| IDH1$_{MS+}$ | T1wp | 0.204 | 0.140 | 1.45 | 0.147 | -0.072 | 0.479 |
| | T1w | -3.211 | 0.562 | -5.71 | $10^{-4}$ | -4.313 | -2.110 |
| | T2w | 1.110 | 0.036 | 30.63 | $10^{-4}$ | 1.039 | 1.182 |
| | FLAIR | 3.673 | 0.182 | 20.16 | $10^{-4}$ | 3.316 | 4.030 |
| | ADC | -0.459 | 0.029 | -15.89 | $10^{-4}$ | -0.516 | -0.402 |
| | cons | -11.176 | 0.487 | -22.93 | $10^{-4}$ | -12.131 | -10.221 |
| MGMT$_{PMS+}$ | T1wp | 2.751 | 0.132 | 20.87 | $10^{-4}$ | 2.493 | 3.009 |
| | T1w | -4.161 | 0.175 | -23.77 | $10^{-4}$ | -4.504 | -3.818 |
| | T2w | 2.275 | 0.069 | 32.76 | $10^{-4}$ | 2.139 | 2.411 |
| | FLAIR | 0.473 | 0.195 | 2.42 | 0.016 | 0.090 | 0.856 |
| | ADC | -4.494 | 0.237 | -18.94 | $10^{-4}$ | -4.959 | -4.029 |
| | cons | -7.075 | 0.270 | -26.21 | $10^{-4}$ | -7.604 | -6.546 |
| CNEC$_+$ | T1wp | -0.798 | 0.020 | -39.91 | $10^{-4}$ | -0.837 | -0.759 |
| | T1w | 4.449 | 0.053 | 84.29 | $10^{-4}$ | 4.345 | 4.552 |
| | T2w | 0.305 | 0.018 | 17.3 | $10^{-4}$ | 0.270 | 0.339 |
| | FLAIR | -2.918 | 0.030 | -98.86 | $10^{-4}$ | -2.976 | -2.860 |
| | ADC | -1.119 | 0.015 | -77.11 | $10^{-4}$ | -1.148 | -1.091 |
| | cons | 1.936 | 0.053 | 36.68 | $10^{-4}$ | 1.832 | 2.039 |
| MVP$_+$ | T1wp | 0.049 | 0.027 | 1.85 | 0.065 | -0.003 | 0.102 |
| | T1w | 1.175 | 0.102 | 11.5 | $10^{-4}$ | 0.975 | 1.375 |
| | T2w | 0.348 | 0.024 | 14.57 | $10^{-4}$ | 0.301 | 0.395 |
| | FLAIR | 1.492 | 0.030 | 50.31 | $10^{-4}$ | 1.434 | 1.550 |
| | ADC | 0.062 | 0.052 | 1.19 | 0.234 | -0.040 | 0.163 |
| | cons | -7.433 | 0.104 | -71.8 | $10^{-4}$ | -7.636 | -7.230 |

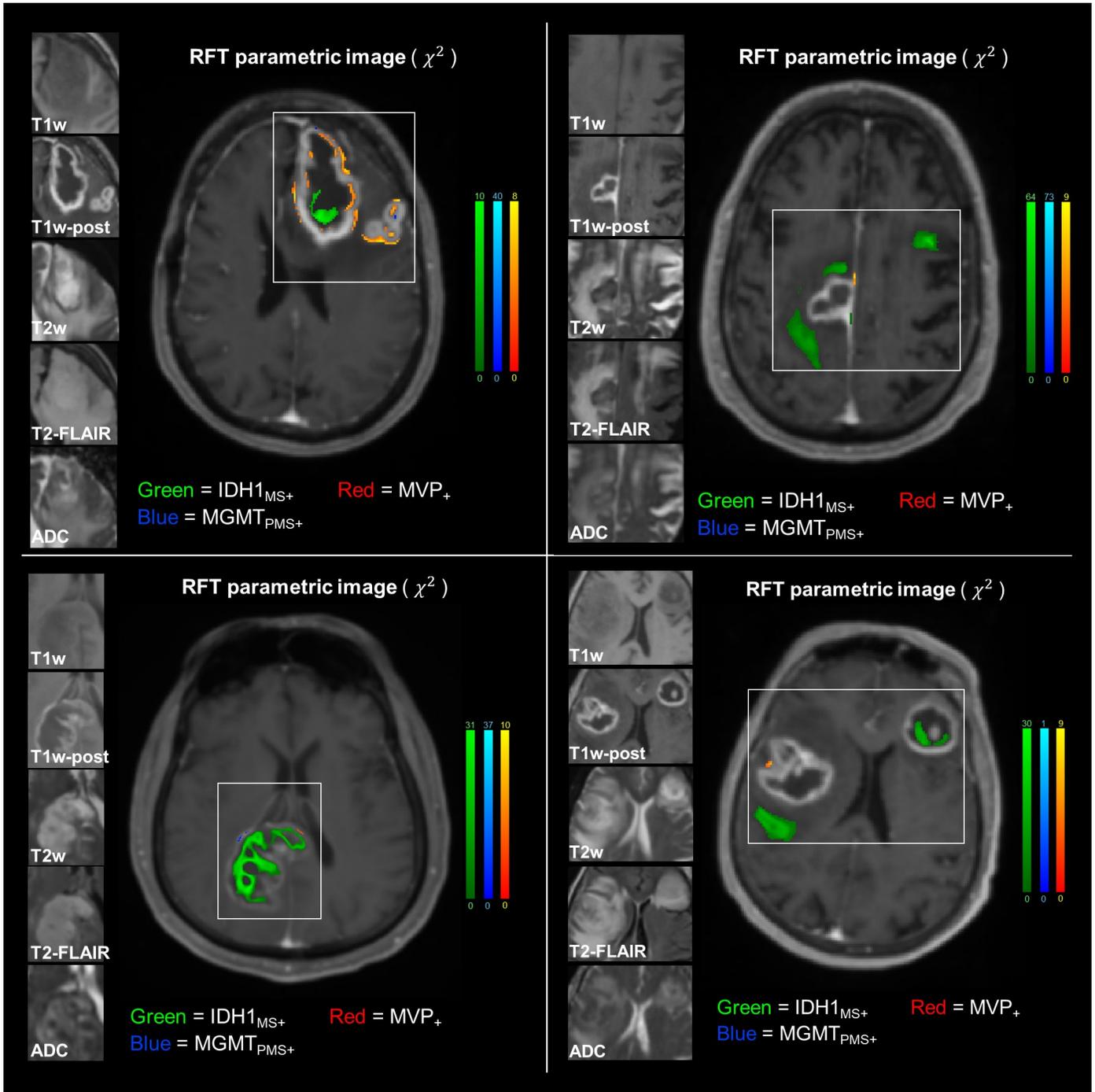

**Supplemental Figure I.** Statistical parametric maps thresholded by RFT for four additional exemplary patients.